\documentclass[aps,prb,twocolumn,psfig,showpacs]{revtex4-1}
\usepackage{amsfonts}
\usepackage{amsmath}
\usepackage{graphicx}
\usepackage{bm}
\usepackage{amssymb}
\usepackage{times}
\usepackage{dcolumn}
\usepackage{cases}
\usepackage{txfonts}
\usepackage{color}

\RequirePackage[hyperindex,colorlinks,bookmarksnumbered,plainpages=true]{hyperref}
\hypersetup{colorlinks,linkcolor=blue,urlcolor=blue,citecolor=blue}
\usepackage{hyperref}

\usepackage{ulem}
\renewcommand{\emph}[1]{\textit{#1}}

\definecolor{darkgreen}{rgb}{0,0.5,0}
\definecolor{darkblue}{rgb}{0,0,0.5}
\definecolor{darkred}{rgb}{.7,0,0}
\definecolor{purple}{rgb}{0.35,0,0.35}
\definecolor{orange}{rgb}{1,0.5,0}
\definecolor{grey}{rgb}{.6,.6,.6}

\makeatletter

\newcommand{\Rmnum}[1]{\expandafter\@slowromancap\romannumeral #1@}
\makeatother

\begin{document}
\title{Fermionic algebraic quantum spin liquid in an octa-kagom\'e frustrated antiferromagnet}
\author{Cheng Peng$^1$, Shi-Ju Ran$^{2}$, Tao Liu$^{1}$, Xi Chen$^1$, and Gang Su$^{1,3}$}
\email[Corresponding author. ]{Email: gsu@ucas.ac.cn}
\affiliation{$^{1}$Theoretical Condensed Matter Physics and Computational Materials Physics Laboratory, School of Physical Sciences, University of Chinese Academy of Sciences, Beijing 100049, China
\linebreak $^{2}$ICFO-Institut de Ciencies Fotoniques, The Barcelona Institute of Science and Technology, 08860 Castelldefels (Barcelona), Spain
\linebreak $^3$Kavli Institute for Theoretical Sciences,  University of Chinese Academy of Sciences, Beijing 100049, China}

\begin{abstract}

We investigate the ground state and finite-temperature properties of the spin-1/2 Heisenberg antiferromagnet on an infinite octa-kagom\'e lattice by utilizing state-of-the-art tensor network-based numerical methods. It is shown that the ground state has a vanishing local magnetization and possesses a $1/2$-magnetization plateau with up-down-up-up spin configuration. A quantum phase transition at the critical coupling ratio $J_{d}/J_{t}=0.6$ is found. When $0<J_{d}/J_{t}<0.6$, the system is in a valence bond state, where an obvious zero-magnetization plateau is observed, implying a gapful spin excitation; when $J_{d}/J_{t}>0.6$, the system exhibits a gapless excitation, in which the dimer-dimer correlation is found decaying in a power law, while the spin-spin and chiral-chiral correlation functions decay exponentially. At the isotropic point ($J_{d}/J_{t}=1$), we unveil that at low temperature ($T$) the specific heat depends linearly on $T$, and the susceptibility tends to a constant for $T\rightarrow 0$, giving rise to a Wilson ratio around unity, implying that the system under interest is a fermionic algebraic quantum spin liquid.

\end{abstract}

\pacs{75.10.Jm, 75.10.Kt, 75.60.Ej, 05.10.Cc}
\maketitle

\section{Introduction}

Quantum spin liquid (QSL), \cite{QSL} also known as quantum disorder or quantum paramagnet, has received considerable attention since it was proposed to describe a possible magnetic disordered state in interacting spin systems even at temperature down to zero. It is intuitive that in two-dimensional (2D) quantum spin models highly geometric frustration and low coordination number usually lead to strong quantum fluctuations, capable of destroying the semi-classical long range orders in the ground state, and thereby inclining to generate a QSL. \cite{QMin2D} In the past decades, there have been extensive numerical simulations \cite{Kagome1-nums,Kagome2-nums,Kagome3-nums,Kagome4-nums,Kagome5-nums} and experimental efforts \cite{Kagome1-exp,Kagome2-exp} showing that the spin-$1/2$ Heisenberg antiferromagnetic model (HAFM) on kagom\'e lattice is the most promising QSL candidate. However, since the intractability of the quantum frustrated system, some unsettled issues are still remaining in hot debate, e.g., whether the ground state of kagom\'e HAFM is a gapped $\mathbb{Z}_2$ spin liquid or a gapless Dirac QSL.

\begin{figure}
    \begin{minipage}{\linewidth}
        \includegraphics[width=\textwidth]{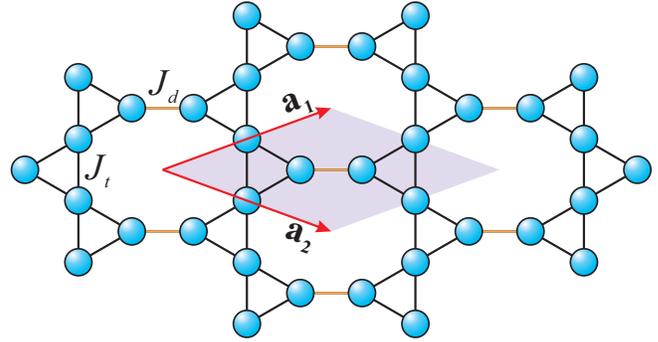}
    \end{minipage}
\caption{\label{fig1}
(Color online) Structure of the octa-kagom\'e lattice (OKL). It can be obtained by stretching the triangles in kagom\'e lattice (KL) along the horizontal direction. If one stretches all three directions of the triangles in KL, it will end up with a star lattice. The OKL can also be viewed as corner and edge sharing octagons. The blue balls represent spins sitting on the lattice site, and the red dashed parallelogram depicts a four-site unit cell, where $\textbf{a}_1$ and $\textbf{a}_2$ are basis vectors. $J_t$ (black) and $J_d$ (orange) denote Heisenberg exchange couplings between nearest neighbor spins inside triangles and inter-triangles, respectively.
}
\end{figure}

Recently, a series of new layered compounds $BiOCu_2(XO_3)(SO_4)(OH)\cdot H_2O$, $(X=Te,Se)$ were discovered \cite{SKM,OKL}. The 2D framework built by magnetic $Cu^{2+}$ ions in these compounds shows an extremely unique lattice (see Fig.~\ref{fig1}). Such a lattice, we dub it as \textit{octa-kagom\'e lattice} (OKL), does not belong to any of the 2D uniform Archimedean lattices, which has not been considered before. OKL can be regarded as a variant of the standard kagom\'e lattice by inserting a dimer between the corner sharing triangles along one direction, which can also be viewed as corner and edge sharing octagons. Owing to strong geometric frustrations and lower coordination numbers in OKL, the spin-1/2 HAFM on OKL could be a long-sought QSL candidate more promising and intriguing than on kagom\'e lattice.

Motivated by the newly synthesized layered compounds $BiOCu_2(XO_3)(SO_4)(OH)\cdot H_2O$, $(X=Te,Se)$, we shall study systematically, for the first time, the ground state and thermodynamic properties of the spin-$1/2$ HAFM on OKL using state-of-the-art tensor network (TN)-based numerical methods. Our results show that the system under investigation possesses a ferminonic algebraic QSL phase. This paper is organized as follows: In Sec. \uppercase\expandafter{\romannumeral2}, the model and TN-based simulating methods are described in detail. In Sec. \uppercase\expandafter{\romannumeral3}, by calculating the local magnetization, we shall show that the ground state of this system is magnetically disordered. In Sec. \uppercase\expandafter{\romannumeral4}, the spatial dependence of spin-spin, dimer-dimer and chiral-chiral correlation functions of the system under interest in the ground state will be given. In Sec. \uppercase\expandafter{\romannumeral5}, the magnetic curves and the phase diagram in the ground state are presented. In Sec. \uppercase\expandafter{\romannumeral6}, the temperature dependence of specific heat and susceptibility will be discussed. Finally, we give a conclusion in Sec. \uppercase\expandafter{\romannumeral7}.

\section{Model and Methods}

The Hamiltonian under interest reads
\begin{equation}\label{eq1}
\textbf{H} = J_{d} \sum\limits_{<ij>}\textbf{S}_{i}\cdot\textbf{S}_{j}+J_{t} \sum\limits_{<lm>}\textbf{S}_{l}\cdot\textbf{S}_{m} -h \sum\limits_{i}S^{z}_{i},
\end{equation}
where $\textbf{S}_i$ is the spin operator on the $i$th site, $J_{d}$ ($J_{t}$) is the coupling constant between nearest neighbor spins standing inside the dimer (triangle), as indicated in Fig.~\ref{fig1}, and $h$ is the magnetic field. We set $J_{t}=1$ as energy scale.

\begin{figure}
    \begin{minipage}{\linewidth}
        \includegraphics[width=\textwidth]{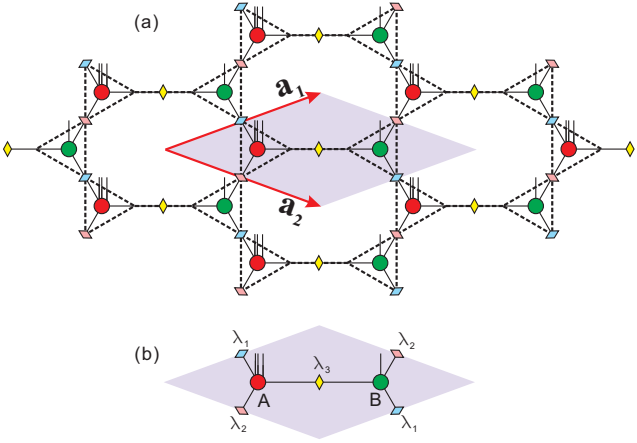}
    \end{minipage}
\caption{\label{fig2}
(Color online) (a) The corresponding ground state TN representation on OKL (black dash). (b) The unit cell containing two non-equivalent tensors $A$ (red circle) and $B$ (green circle), and three different diagonal matrices $\lambda_1$ (blue diamond), $\lambda_2$ (pink diamond) and $\lambda_3$ (yellow diamond).
}
\end{figure}

\begin{figure}
    \begin{minipage}{\linewidth}
        \includegraphics[width=\textwidth]{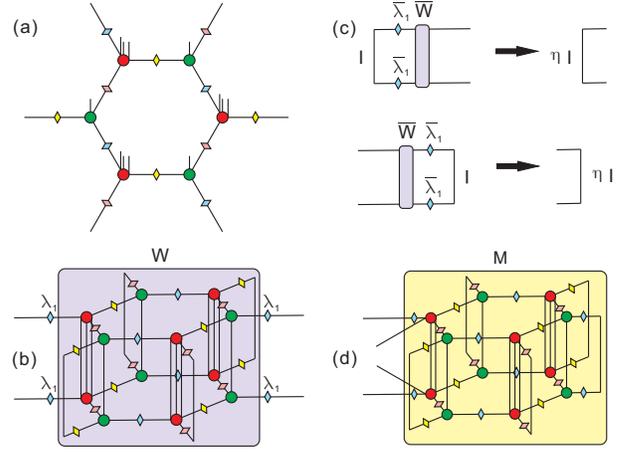}
    \end{minipage}
\caption{\label{fig3}
(Color online) Graphical representation for the cluster update scheme. (a) The hexagon cluster is composed of six tensors and twelve diagonal matrices. (b) We contract the physical indices and virtual bonds connected with diagonal matrices $\lambda_2$ and $\lambda_3$ of double-layer tensor cluster to get tensor W, which will later be canonicalized for updating the diagonal matrices $\lambda_1$, as well as the tensors A and B. (c) The orthogonal conditions for renewed $\overline{\lambda}_1$ and $\overline{W}$. (d) Extracting a physical index from the hexagon cluster by Tucker decomposition.
}
\end{figure}

\begin{figure}
    \begin{minipage}{0.7\linewidth}
        \includegraphics[width=\textwidth]{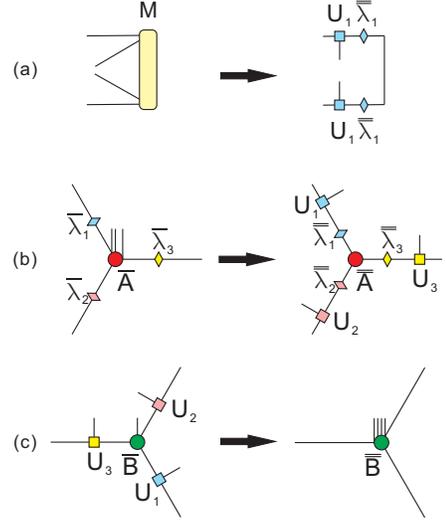}
    \end{minipage}
\caption{\label{fig4}
(Color online) (a) SVD decomposition of the double layer cluster in Fig.~\ref{fig3} (d). (b) Permutation of physical indices from $A$ using Tucker decomposition. (c) Absorbing the physical indices into $B$.
}
\end{figure}

\begin{figure}
    \begin{minipage}{0.7\linewidth}
        \includegraphics[width=\textwidth]{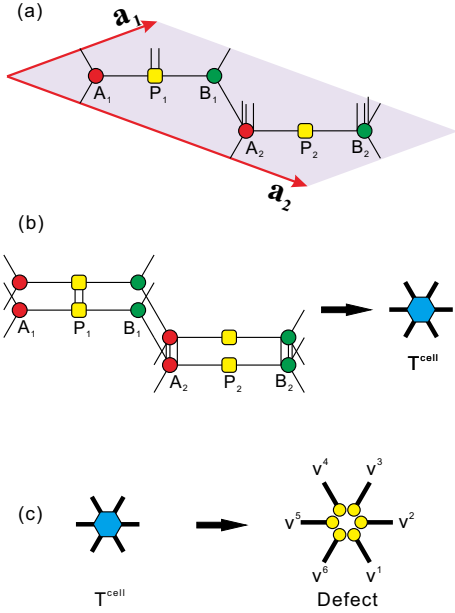}
    \end{minipage}
\caption{\label{fig5}
(Color online) (a) Unit cell of TN in NCD scheme, where $\textbf{a}_1$ and $\textbf{a}_2$ are basis vectors. (b) The construction of $T^{cell}$. (c) Transformation from $T^{cell}$ to ``defect''. The ``defect'' is constructed by six contractors $v^n$ ($n=1, 2, \ldots, 6$) obtained by Eq.~(\ref{eq6}), and each contractor is denoted by a yellow circle connected with a bold black line.
}
\end{figure}

It is usually challenging to simulate quantum many-body systems. Due to strong correlations and quantum fluctuations, most traditional methods fail to capture their novel properties. For example, mean-field theories underestimate long range fluctuations that may be critically important to the exotic many-body phenomena; quantum Monte Carlo suffers from the notorious sign-problem \cite{QMCsign} when calculating frustrated spin models as well as fermionic models away from the half filling; etc.

In this work, we use state-of-the-art TN algorithms to explore the spin-1/2 Heisenberg antiferromagnet on OKL. TN scheme is free from negative-sign problem, and has been demonstrated to be a powerful numerical tool not only in strongly correlated quantum systems, \cite{QMBSDMRGPEPS,QMBS1,QMBS2} but also in statistical physics, \cite{statistic1-PEPS,statistic2-TRG,statistic3-HOSRG,statistic4} quantum information  \cite{QuantumInformation1,QuantumInformation2,QuantumInformation3} and so on. The central task in such kind of algorithms is to compute the TN contraction, \cite{QuantumInformation2,TNComp} i.e. to sum over all shared bonds in TN. However, except some special cases, \cite{Tree,MERA,exactcontrac1,exactcontrac2} the contraction of the TN with a regular geometry (e.g. square or honeycomb) has been shown to be NP-hard. \cite{QuantumInformation2}

Generally, there are two ways to deal with the TN simulations: renormalization \cite{statistic2-TRG,iTEBD,statistic1-PEPS,PEPS1,PEPS2,iPEPS,TERG,CTMRG1,CTMRG2,SRG1,SRG2,statistic3-HOSRG} and encoding \cite{ODTNS,NCD,AOP} schemes. The former follows a contraction-and-truncation scheme, while the latter encodes the TN contraction into a local self-consistent problem. Specifically, the renormalization scheme originates from Wilson's numerical renormalization group method, \cite{NRG,NRG-Kondo} which solves successfully the Kondo \cite{Kondo} problem. Then, the density matrix renormalization group (DMRG) \cite{DMRG1,DMRG2} was proposed by White, where the boundary condition (especially in 1D) is better considered with entanglement. For 2D systems, algorithms based on tensor renormalization group and the infinite projected entangled pair state (iPEPS) \cite{iPEPS} were proposed. The degrees of freedom is coarse-grained in such a way that when the tensor is invariant under renormalization, it represents approximately an infinite system.

The encoding scheme follows an opposite way known as the ``mean-field'' idea that considers well the entanglement with the help of TN. The ``mean-field'' idea is incredibly important in numerical physics, which gives birth of the great density functional theory \cite{DFTreview1,DFTreview2} and \textit{ab-initio} scheme which has been widely used in both physics and chemistry. To better deal with the strong correlations in many-body physics, the dynamic mean-field theory \cite{DMFT1,DMFT2,DMFT3,DMFT4,DMFT5} and density matrix embedding theory \cite{DMET1,DMET2,DMET3} were also proposed. By combining ``mean-field'' idea with TN and multi-linear algebra, the \textit{ab-initio} optimization principle was proposed, \cite{AOP} where an infinite TN is equivalently transformed into a local tensor embedded in an entanglement bath.

We here employ three kinds of TN-based algorithms, namely cluster update \cite{cluster1,cluster2,cluster3} and full update schemes \cite{iPEPS,full-update2,full-update3} of the iPEPS \cite{PEPS1,PEPS2,iPEPS} (a contraction-and-truncation scheme) and network contractor dynamics \cite{NCD} (NCD) approach (an encoding scheme) to investigate our model for mutually validating the results obtained by each scheme. Consequently, the calculated results are consistent with each other, which manifests itself the reliability of our simulations.

\subsection{Tensor-network state ansatz}

We start from a TN state ansatz, as shown in Fig.~\ref{fig2}, to describe the states at zero temperature. Such a TN state is composed of two non-equivalent tensors $A$ and $B$, and three different diagonal matrices $\lambda_1$, $\lambda_2$ and $\lambda_3$. $A$ and $B$ (each of which contains three virtual bonds that carry the entanglement of the state) are located on the two inequivalent triangles of the OKL, respectively. The physical degrees of freedom of the three spins in triangle A are put on tensor $A$, where the dimension of the physical space is 8. In this way, the dimension of the physical bond of tensor $B$ is 2, which is the Hilbert space of the spin on the right side with the $J_{d}$ coupling. Mathematically, such a TN state is written as

\begin{widetext}
\begin{equation}\label{eq2}
|\Psi\rangle = \sum_{\{s\}} \sum_{\mu, \nu, \xi=1}^{D_c} \Bigg( \prod_{k\in\it TN} [\lambda_{1}]_{\mu \mu}[\lambda_{2}]_{\nu \nu}[\lambda_{3}]_{\xi \xi} A_{\mu \nu \xi}^{s_{k,1}s_{k,2}s_{k,3}} B_{\mu \nu \xi}^{s_{k,4}} |s_{k,1}s_{k,2}s_{k,3}s_{k,4}\rangle \Bigg),
\end{equation}
\end{widetext}
where $k$ refers to the $k$-th unit cell of the whole lattice with $\textbf{a}_1$ and $\textbf{a}_2$ basis vectors [see Fig.~\ref{fig2} (b)]. To get the ground state, the imaginary time evolution is implemented to minimize the energy of the PEPS by

\begin{equation}\label{eq3}
  |\Psi_{gs}\rangle = \lim_{\beta \rightarrow \infty} \frac{e^{-\beta \textbf{H}}|\Psi\rangle}{\parallel{e^{-\beta \textbf{H}}|\Psi\rangle}\parallel},
\end{equation}
where $\beta=1/{k_B T}$.

It is impossible to calculate Eq.~(\ref{eq3}) exactly in the thermodynamic limit, since the dimension of $\textbf{H}$ increases exponentially with the number of lattice sites. Here, we use the Trotter-Suzuki decomposition to implement the evolution on the TN state. By splitting Hamiltonian into two parts, one has $H_{a}=\sum_{k} H_{left-trangle}^{[k]}$ and $H_{b}=\sum_{k} \bigg( H_{dimer}^{[k]}+H_{right-trangle}^{[k]}\bigg)$, and the first-order Trotter-Suzuki decomposition can be used to approximate the evolution operator, i.e., $e^{-\beta H} \approx \big(e^{\tau H_a}e^{\tau H_b}\big)^N+\mathcal{O}(\tau^2)$, with $\beta=N\tau$. The approximation becomes accurate when the Trotter step $\tau$ approaches zero. In practical calculations, we decrease $\tau$ gradually from $1\times 10^{-1}$ to $1\times 10^{-5}$ so that the Trotter error becomes negligible.

By considering the translation invariance, we can adopt the local operation instead of evolving the whole system, and optimize the environment around the local tensors. Incidentally, for finite-temperature thermal states, the imaginary-time evolution of the density operator can be implemented similarly.

\subsection{Cluster update}

We choose a hexagon consisting of six tensors as the environment for cluster update, as depicted in Fig.~\ref{fig3} (a). The cluster tensors are transformed into a super-orthogonal form \cite{ODTNS} in order to approximate the global environment optimally. Taking Fig.~\ref{fig3} (b) as an example, we build a double-layer structure of the cluster tensor and contract all physical indices and virtual bonds on the \textit{bra} and \textit{ket} layers except the bonds connected by $\lambda_1$. For convenience, the shaded part of Fig.~\ref{fig3} (b) is represented by $W$. The super-orthogonalization is much like the canonicalization for an infinite 1D lattice \cite{iTEBD-canon}. The update of $\overline{\lambda}_1$ and $\overline{W}$ leads to the conditions

\begin{equation}\label{eq4}
  \sum_{\mu,\mu'=1}^{D_c}\delta_{\mu \mu'}[\overline{\lambda}_{1}]_{\mu \mu}[\overline{\lambda}_{1}]_{\mu' \mu'}\overline{W}_{\mu \mu', \nu \nu'}=\eta \delta_{\nu \nu'},
\end{equation}

\begin{equation}\label{eq5}
  \sum_{\nu,\nu'=1}^{D_c}\overline{W}_{\mu \mu', \nu \nu'}[\overline{\lambda}_{1}]_{\nu \nu}[\overline{\lambda}_{1}]_{\nu' \nu'}\delta_{\nu \nu'}=\eta \delta_{\mu \mu'}.
\end{equation}

Fig.~\ref{fig3} (c) is the graphical representation of Eqs. (\ref{eq4}) and (\ref{eq5}). The update of $W$ is actually acting on $A$ and $B$ along the $\lambda_{1}$ direction, where $A$ and $B$ are renewed to $\overline{A}$ and $\overline{B}$. Operations on the other two directions are similar. We iterate this procedure until the cluster satisfies simultaneously the orthogonality conditions in all three directions. Then, the environment of the cluster can be best approximated by the converged diagonal matrices $\overline{\lambda}_1$, $\overline{\lambda}_2$ and $\overline{\lambda}_3$.

Then we permute the physical indices from A to B to evolve the interactions on the B triangles. This operation will increase the bond dimensions, and a truncation is needed. Taking Fig.~\ref{fig3} (d) as an example, we leave one physical index and the corresponding virtual bonds of A open and others contracted in the cluster. We use $M$ to denote the intermediate reduced density matrix, where the dimension of $M$ is $2D_c\times 2D_c$. Moreover, $M$ is a Hermitian matrix because of the double-layer structure. Then, we decompose $M$ using the SVD and only keep the basis corresponding to the $D_c$ dominant singular values. This procedure is shown in Fig.~\ref{fig4} (a), where $U_1$ is the unitary matrix given by the SVD holding the spared physical index of $M$, and $\overset{=}{\lambda}_1$ is the square root of the singular spectrum after truncation. $U_2$, $U_3$, $\overset{=}{\lambda}_2$ and $\overset{=}{\lambda}_3$ are obtained in the similar way.

Finally, we change the position of all three physical indices from $\overline{A}$ into $\overline{B}$, as depicted in Figs.~\ref{fig4} (b) and (c). In such a way, the evolutions given by the interactions of the triangles A and B are implemented in turn, where the geometry and the bond dimensions are kept unchanged.

\subsection{Full update}

Unlike the cluster update scheme, the full update scheme needs to contract all the 2D TN in order to truncate and obtain physical quantities. There are two widely used ways to simulate the whole environment, namely the iTEBD \cite{iTEBD} and the corner transfer matrix renormalization group (CTMRG) \cite{CTMRG1,CTMRG2}. Here, we use the iTEBD in our calculation, where the TN is contracted to a matrix product state (MPS) on its boundary. Full update can achieve a higher accuracy than local optimization methods, but the computation cost is significantly large. We set the bond dimension of the MPS in iTEBD as $D_c^2$ to balance the accuracy and cost. The permutation of physical indices and variational optimization of the truncation matrices follow Refs.~\onlinecite{iPEPS,cluster3,full-update2,full-update3}.

\subsection{Network Contractor Dynamics}

NCD was first proposed to solve the TN contractions in the calculation of partition function of 2D quantum models. We adopt the NCD algorithm to optimize the environment around the local tensors in ground state simulation. Different from the renormalization, NCD follows a TN encoding strategy \cite{AOP}, and the TN structure is also different from that in Fig.~\ref{fig2}. The specific cell tensor is shown in Fig.~\ref{fig5} (a), which contains $6$ nonequivalent tensors with $A_1$, $B_1$, $A_2$ and $B_2$ located on triangles, and $P_1$, $P_2$ on dimers. When mapping onto the OKL, there are eight inequivalent lattice sites in the cell tensor of NCD, twice as large as the cell tensors of cluster update and full update schemes. Physical indices are on $P_1$, $A_2$ and $B_2$, indicating that the Hamiltonian splits into two parts, where the first part contains interactions between spins sitting on the dimer denoted by $P_1$ and triangles denoted by $A_2$ and $B_2$, and the rest of interactions are included in the second part of the Hamiltonian. Imaginary time evolution is applied to minimize the ground state energy, and consequently, NCD procedure plays the role of super-orthogonalization to approximate the whole TN contraction.  As explained in Ref.~\onlinecite{NCD}, the whole TN contraction is simplified to a local contraction of a tensor cluster $T^{cell}$ with six contractors ${v^n},n \in [1,2,3,4,5,6]$. $T^{cell}$ is a six-order tensor obtained by contracting the physical indices and connected virtual bonds of the double-layer cell tensors, as represented in Fig.~\ref{fig5} (b), where we use bold black lines to indicate a fat index that contains double virtual bonds in one of the six directions. While each contractor $v^n$ is a vector with the same dimension as the $n$th index of $T^{cell}$. $T^{cell}$ and ${v^n}$ need to meet the following self-consistent relation

\begin{equation}\label{eq6}
\sum_{g_{n \neq \varepsilon}}{T^{cell}_{g_1g_2g_3g_4g_5g_6} \prod_{n \neq \varepsilon}{v^n_{g_n}}} \propto v^{\varepsilon}_{g_{\varepsilon}}.
\end{equation}

Eq.~(\ref{eq6}) includes six self-consistent equations for $\varepsilon \in [1,2,3,4,5,6]$, which should be satisfied simultaneously. Analytically, the six contractors solved from Eq.~(\ref{eq6}) are precisely those given by rank-1 decomposition of $T^{cell}$. The rank-1 tensor, which we call a ``defect'', is given by a direct product of the six contractors. The graphic representation of ``defect'' is shown in Fig.~\ref{fig5} (c). The ``defect'' is actually the first-order approximation of $T^{cell}$. If one substitutes the minimal number of $T^{cell}$'s with the ``defects'' so that no loop appears, then the original TN will become a tree framework. Thanks to the self-consistent conditions, there is no need to compute the whole contraction of such a tree, and only are the local contraction of $T^{cell}$ and the contractors ${v_n}$ required. In this sense, the physical quantities calculated from the ``defective'' TN can be viewed as a mean-field approximation of the exact one. In addition, we can introduce more loops into the defected TN to achieve a higher accuracy, but the computing cost increases inevitably.

\section{Disordered ground state}

\begin{figure}
    \begin{minipage}{\linewidth}
        \includegraphics[width=\textwidth]{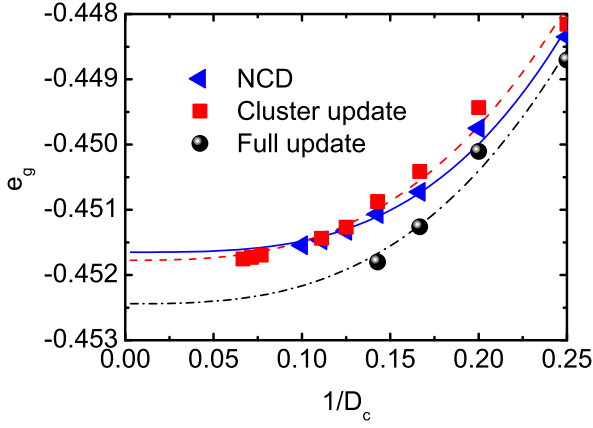}
    \end{minipage}
\caption{\label{fig6}
(Color online) The calculated energy per site for ground state $e_g$ versus inverse bond dimension $1/D_c$ (up to $D_c=10$ for NCD, $D_c=15$ for cluster update, and $D_c=7$ for full update). It is shown that $e_g$ decreases with enhancing $D_c$. Power law fittings for NCD (blue solid line), cluster update (red dashed line) and full update (black dash-dote line) are given, and $e_g$ is extrapolated to be -0.4517 (NCD), -0.4518 (cluster update), and -0.4524 (full update) in the infinite $D_c$ limit.
}
\end{figure}

\begin{figure}
    \begin{minipage}{\linewidth}
        \includegraphics[width=\textwidth]{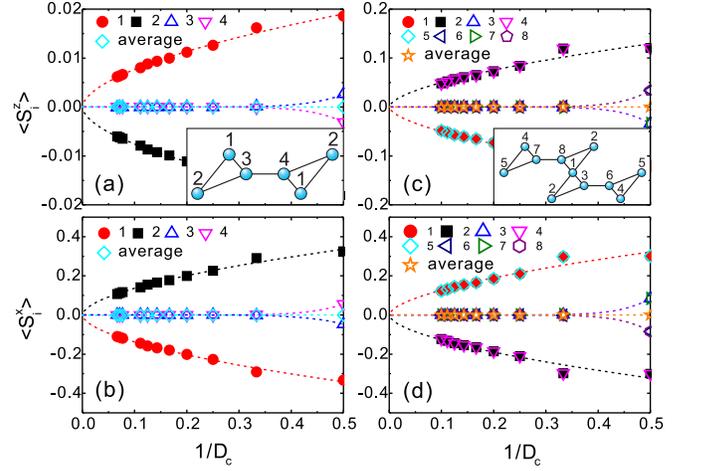}
    \end{minipage}
\caption{\label{fig7}
(Color online)  The local magnetic moment (a) $\langle S^{z}_{i}\rangle$ and (b) $\langle S^{x}_{i}\rangle$ versus inverse bond dimension $1/D_c$ calculated by cluster-update, where $i=1,2,3,4$ denote four inequivalent sites in a unit cell marked in the inset of (a), and the ``average'' represents the intracellular mean magnetic moment. The local magnetic moment (c) $\langle S^{z}_{i}\rangle$ and (d) $\langle S^{x}_{i}\rangle$ versus inverse bond dimension $1/D_c$ calculated by NCD, where $i=1,2,3,4,5,6,7,8$ denote eight sites in an expanded cell in the inset of (c), and the ``average'' denotes the average of the magnetic moments over these eight sites.
}
\end{figure}

 Let us now study the ground state properties of the spin-1/2 HAFM on the infinite OKL for isotropic point ($J_{d}=1$). To testify the reliability of our calculations, we compare the ground state energy $e_g$ (per site) obtained by different methods including NCD, cluster and full updates of iPEPS (Fig.~\ref{fig6}). We found that for large bond dimension $D_c>4$, all schemes give consistent results, showing the reliability of our calculations. A power law dependence is found, with which the energy of infinite $D_c$ is $e_g=-0.4524$ by extrapolation, which is lower than $-0.4386(5)$, the extrapolated ground state energy of the spin-1/2 HAFM on kagom\'e lattice given by DMRG. \cite{Kagome3-nums,Kagome4-nums}

In Fig.~\ref{fig7}, we present the local magnetization $\langle S_{i}^{\alpha}\rangle$ on each nonequivalent site in a unit cell for $h=0$. Small values of $\langle S_{i}^{z}\rangle$ and $\langle S_{i}^{x}\rangle$ caused by the truncation error can be observed. By increasing $D_c$, $\langle S_{i}^{z}\rangle$ and $\langle S_{i}^{x}\rangle$ decay rapidly, and the data are fitted (the dashed lines in Fig.~\ref{fig7}) with the function $f(D_c)=p(1/D_c)^q$, where $p$ and $q$ are fitting parameters. Thus, the extrapolation of local magnetic moments gives a zero magnetization in $D_c \rightarrow \infty$ limit. In particular, as the average values of $\langle S_{i}^{z}\rangle$ and $\langle S_{i}^{x}\rangle$ only fluctuate in the vicinity of zero with the increase of $D_c$, we may use a linear fitting for the average magnetic moments that gives negligible intercepts about $10^{-5}$. The absence of local magnetic moments strongly suggest that it does not have conventional magnetic orders in the ground state, i.e., no traditional SO(3) symmetry is broken.

\section{Spin-spin, dimer-dimer and chiral-chiral correlation functions}

\begin{figure}
    \begin{minipage}{\linewidth}
        \includegraphics[width=\textwidth]{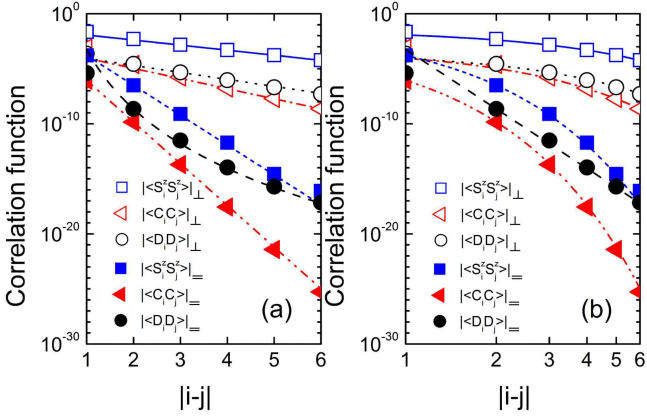}
    \end{minipage}
\caption{\label{fig8}
(Color online) Spatial dependence of correlation functions of the spin-1/2 HAFM on OKL in (a) semi-log and (b) log-log plots. The vertical direction is represented by subscript``$\perp$" and the horizontal direction is marked by``$=$". In vertical direction, the spin-spin (fitting with blue solid line), chiral-chiral (fitting with red dash-dot line) and dimer-dimer (fitting with black dot line) correlation functions show exponential decaying behaviors. In horizontal direction, the spin-spin (fitting with blue short-dash line), chiral-chiral (fitting with red dash-dot-dot line) correlation functions show exponential decaying behaviors, while the dimer-dimer (fitting with black dashed line) correlation function shows a power law decay. All correlation functions are calculated by the full update algorithm with $D_c=5$.
}
\end{figure}

In Fig.~\ref{fig8}, we present the spatial dependence of several correlation functions in the ground state for the system under interest with $J_{d}=1$. The spin-spin correlation function $|\langle S_{i}^{z} S_{j}^{z}\rangle|$ along the horizontal axis is found to decay exponentially, satisfying $f(|i-j|)=\alpha \exp(-|i-j|/\xi)$ with $\alpha=0.084$ and the correlation length $\xi=0.16$, which shows that the spin-spin correlation of this system is short-ranged and the ground state is magnetically disordered.

The chiral-chiral correlation function is defined by $|\langle C_{i} C_{j}\rangle|=|\langle [\textbf{S}_{i_{1}}\cdot (\textbf{S}_{i_{2}}\times \textbf{S}_{i_{3}})] [\textbf{S}_{j_{1}} \cdot (\textbf{S}_{j_{2}}\times \textbf{S}_{j_{3}})]\rangle-\langle \textbf{S}_{i_{1}}\cdot (\textbf{S}_{i_{2}}\times \textbf{S}_{i_{3}})\rangle \langle\textbf{S}_{j_{1}} \cdot (\textbf{S}_{j_{2}}\times \textbf{S}_{j_{3}})\rangle|$, where the lattice sites $i$ and $j$ belong to the left-triangles along the horizontal direction. It is found that the chiral-chiral correlation function also decays exponentially with $\alpha=0.0054$ and $\xi=0.11$, revealing the absence of a long-range spin chiral order.

The dimer-dimer correlation function, which is defined by $|\langle D_{i} D_{j}\rangle|=|\langle (S_{i}^{z} S_{i+1}^{z})(S_{j}^{z} S_{j+1}^{z})\rangle-\langle S_{i}^{z} S_{i+1}^{z}\rangle \langle S_{j}^{z} S_{j+1}^{z}\rangle|$ for the $i$-th and $j$-th dimers, is disclosed to exhibit a power-law decay as of the form $|\langle D_{i} D_{j}\rangle|\sim 1/|i-j|^{\eta}$ with $\eta=17.96$ [Fig.~\ref{fig8}(b)]. This fact signatures possible existence of an algebraic QSL in this system.

Here it is interesting to ask if the correlations along the vertical axis behave the same as those along the horizontal axis. To answer this question, we also calculated the three correlation functions in the vertical direction. The results show that the behaviors in this direction are different, as shown in Fig. ~\ref{fig8}. It is seen that all three correlations along the vertical axis decay exponentially, fitted by the function $f(|i-j|)=\alpha \exp(-|i-j|/\xi)$, with $|\langle S_{i}^{z} S_{j}^{z}\rangle|$ fitted with $\alpha=0.034$ and $\xi=0.94$, $|\langle C_{i} C_{j}\rangle|$ fitted with $\alpha=0.00075$ and $\xi=0.47$, and $|\langle D_{i} D_{j}\rangle|$ fitted with $\alpha=0.00038$ and $\xi=0.68$. This is owing to the nonequivalent lattice structure along the two axes. It is the introduction of $J_d$ that causes the lattice essentially distinct from a combination of decoupled zig-zag spin chains, of which the ground state is a valence bond state (VBS) with two-fold degeneracy and a finite magnetic excitation gap. \cite{sawtooth1,sawtooth2} A strong $J_d$ coupling (especially at the isotropic point) is crucial in the critical phase, which will be discussed later.

We would like to mention that the nature of the correlations presented here is similar to the case with a resonating valence bond (RVB) wave function constructed on a square lattice, where an exponentially decaying spin-spin correlation and a power-law decaying dimer-dimer correlation were also observed. \cite{NNRVB square,J1-J2 square}

\section{Magnetic curves and phase diagram in ground state}

\begin{figure}
    \begin{minipage}{\linewidth}
        \includegraphics[width=\textwidth]{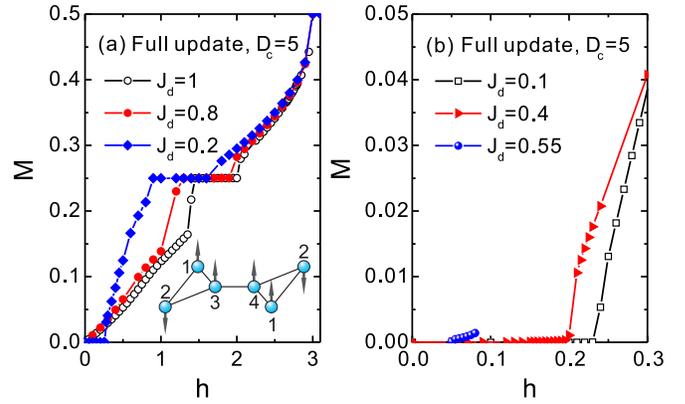}
    \end{minipage}
\caption{\label{fig9}
(Color online) Magnetic curves of the spin-1/2 HAFM on OKL. (a) $J_{d}=0.2,0.8$ and $1$, (b) $J_{d}=0.1,0.4$ and $0.55$ under low magnetic fields. Inset of (a): Up-down-up-up (UDUU) spin configuration in a unit cell in the $M=1/4$ plateau phase.
}
\end{figure}

\begin{figure}
    \begin{minipage}{\linewidth}
        \includegraphics[width=\textwidth]{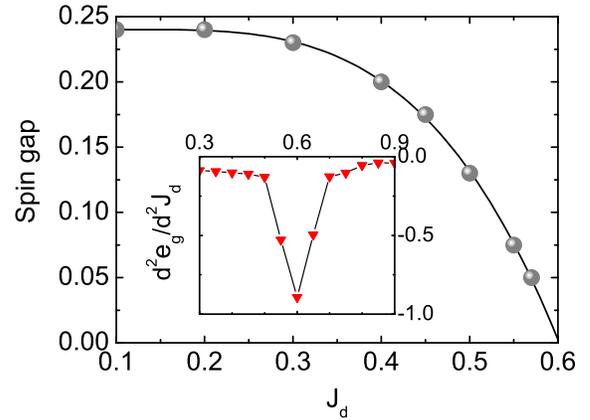}
    \end{minipage}
\caption{\label{fig10}
(Color online) Spin gap as a function of $J_{d}$ for the spin-1/2 HAFM on OKL. The inset is the second-order derivative of the ground state energy  with respect to $J_{d}$ in the absence of a magnetic field. It is clear that the point $J_d= 0.6$ is singular, at which the spin gap closes.
}
\end{figure}

\begin{figure}
    \begin{minipage}{0.8\linewidth}
        \includegraphics[width=\textwidth]{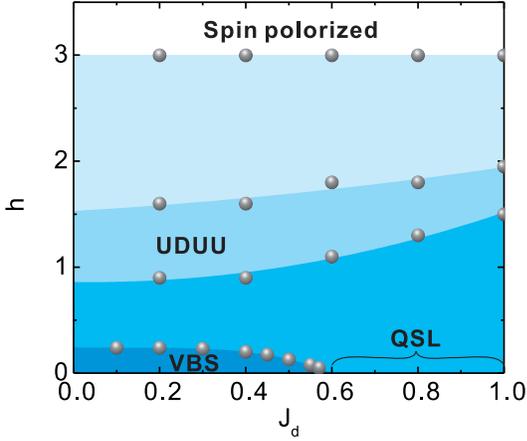}
    \end{minipage}
\caption{\label{fig11}
The ground state phase diagram for the spin-1/2 HAFM on OKL. A quantum phase transition from VBS phase to QSL phase happens at the critical point $J_d^c= 0.6$. UDUU: the 1/2-magnetization plateau phase with up-down-up-up spin configuration.
}
\end{figure}

The magnetization per site $M$ as a function of magnetic field $h$ in the spin-1/2 HAFM on OKL is presented in Fig.~\ref{fig9}. One may observe that in magnetic curves [Fig. \ref{fig9} (a)], for $J_{d}=0.2$, three plateaux with $M=0, \frac{1}{4}$ and $\frac{1}{2}$ are observed, while for $J_{d}=0.8$ and $1$, apart from the two plateaux with $M=\frac{1}{4}$ and $\frac{1}{2}$, no $M=0$ plateau is found. These results imply that depending on $J_d$, there may be two phases in the system, one phase with a zero-magnetization plateau and the other phase without. As the width of the $M=0$ plateau gives the gap from the singlet ground state to the first triplet excited state, we find that in the phase with small $J_{d}$ the spin excitation is gapful, while in the other phase with large $J_{d}$ it is gapless. For a closer inspection, we calculated the cases with small $J_{d}$ under weaker magnetic fields, as given in Fig.~\ref{fig9} (b). The results demonstrate that the spin gap decreases with increasing $J_{d}$, suggesting that there must be a critical point $J_d^c$, at which a quantum phase transition (QPT) happens: for $J_d<J_d^c$ the ground state is in a gapped phase, and for $J_d>J_d^c$ it is in a gapless phase.

Another interesting phenomenon in magnetic curves is the occurrence of $M=1/4$-plateau, which can also be called $M/M_s=1/2$-plateau (briefly 1/2-magnetization plateau) with $M_s=1/2$ the saturation magnetization per site. The frustrated Heisenberg models on lattices with triangular structures lead usually to 1/3-magnetization plateau, which has been found in, e.g., kagom\'e \cite{kagomeplateau1,kagomeplateau2,kagomeplateau3} and Husimi \cite{Husimi} lattices. The occurrence of the 1/2-magnetization plateau in the present system is understandable, because the unit cell of the OKL contains four inequivalent lattice sites, leading to the periodicity $n$ of the ground state is 4, consistent with the condition of $n(S-M)=integer$. To explore the nature of this 1/2-plateau, we calculated the local magnetic moment at four inequivalent sites in a unit cell, and found in this plateau phase the spin configuration is of up-down-up-up (UDUU), as illustrated in the inset of  Fig.~\ref{fig9} (a). Such a plateau is a commensurate, classical state stabilized by quantum fluctuations.

To determine accurately the quantum critical point (QCP) $J_d^c$, we calculated the spin gap as a function of $J_d$, as given in Fig.~\ref{fig10}, which gives $J_d^c=0.6$. To further confirm this point, we also studied the second-order derivative of the ground state energy with respect to $J_d$ (the inset of Fig.~\ref{fig10}), which reveals a sharp dip at the same point, indicating the QPT indeed appears at $J_d^c=0.6$.

By summarizing our calculated results, we present the ground state phase diagram of the spin-1/2 HAFM on OKL in the $J_d-h$ plane, as shown in Fig.~\ref{fig11}. It can be seen that when $h=0$, the phase for $J_d<0.6$ is a VBS, as in the limit of $J_d \rightarrow 0$, the system approaches to an uncoupled zig-zag spin chain, whose ground state is a VBS with twofold degeneracy and a finite magnetic excitation gap. \cite{sawtooth1,sawtooth2} Because there is no quantum phase transition  for $J_d<J_d^c$, the system should stay in the same VBS phase. For $J_d>0.6$, the system enters into a gapless QSL state, which is evidenced by the algebraically decaying dimer-dimer correlations and vanishing local magnetic moments. When $h>0$, the VBS state is gradually melted by closing the spin gap, and the system enters into a spin canted state.  By increasing the magnetic field further, the system enters into the 1/2-magnetization plateau (with $M=1/4$) phase, in which the spin gap opens again, and the spin configurations are arranged in UDUU alignments. Above the UDUU phase, it enters into another spin canted phase. By increasing $h$ further, all spins are polarized. It should be remarked that all the phase boundaries in this phase diagram are obtained by observing various critical magnetic fields.

\section{Thermodynamic properties}

\begin{figure}
    \begin{minipage}{\linewidth}
        \includegraphics[width=\textwidth]{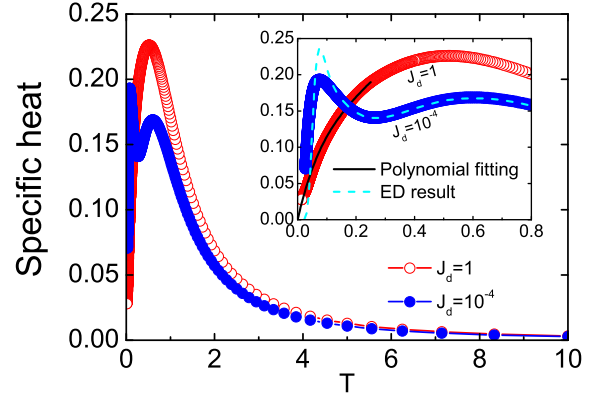}
    \end{minipage}
\caption{\label{fig12}
(Color online) Specific heat $C(T)$ versus temperature $T$ of the spin-$1/2$ HAFM on OKL for $J_{d}=10^{-4}$ (blue solid circle) and $J_{d}=1$ (red open circle).  Inset: the low-temperature part of $J_{d}=1$ below $T=0.25$, which can be well fitted by a polynomial $C(T)=1.741T-7.96T^2+22.51T^3-25.61T^4$ (black line), and that of $J_{d}=10^{-4}$,  which is also better compared with an exact diagonalization (ED) result (cyan dash) of the zig-zag spin chain containing 8 triangles. Here the bond dimension is $D_c=20$.
}
\end{figure}

\begin{figure}
    \begin{minipage}{\linewidth}
        \includegraphics[width=\textwidth]{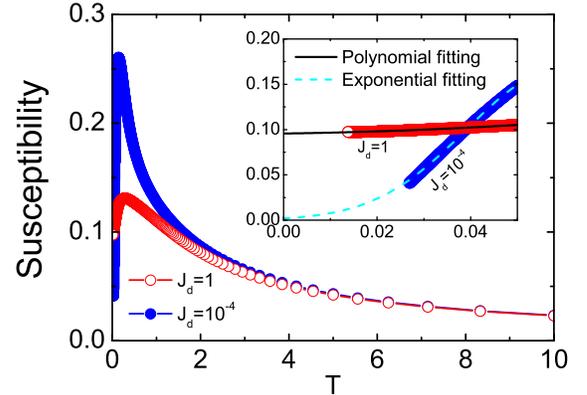}
    \end{minipage}
\caption{\label{fig13}
(Color online) Zero-field magnetic susceptibility $\chi$ as a function of temperature $T$ of the spin-$1/2$ HAFM on OKL for $J_{d}=1$ (red open circle) and $J_{d}=10^{-4}$ (blue solid circle). Inset: the low-temperature part of susceptibility, where  the case of $J_{d}=1$ can be fitted with a polynomial $\chi(T)=0.09566-0.06828T+2.438T^2$ (black line), and that of $J_{d}=10^{-4}$ behaviors in an exponential way. Here the bond dimension is $D_c=20$.
}
\end{figure}

Next, we explore the thermodynamic properties of the spin-$1/2$ HAFM on OKL using the optimized decimation of tensor network state. \cite{ODTNS} The free energy can be obtained by collecting all renormalization factors down to the targeted temperature. Alternatively, one can also get the physical quantities by calculating the expectation values of local operators with tensor-network thermodynamic states. Considering the precision and cost of thermal-state TN algorithms, we choose the cluster update scheme to contract the ``environment" around the local inequivalent tensors. The energy as well as other thermodynamic quantities including specific heat and susceptibility are thus calculated. To keep a higher accuracy, we adopted the second-order Trotter-Suzuki decomposition and fix the Trotter slice to be 0.01 in the calculations of thermodynamic properties.

We  obtain the temperature dependence of the specific heat by $C(T)={\partial f}/{\partial T}$, where $f$ is the free energy per site. Fig.~\ref{fig12} gives the results for $J_{d}=10^{-4}$ and $1$. It is observed that at high temperature, both go to converge, and $C(T)$ decreases down to zero with increasing temperature. But at low temperature (see the inset of Fig.~\ref{fig12}), both cases show intrinsically distinct behaviors: the specific heat for $J_{d}=10^{-4}$ exhibits two peaks and is pretty close to the exact diagonalization (ED) result of the zig-zag spin chain with 8 triangles, which also verifies the reliability of our method. When $T\rightarrow 0$, $C(T)$ shows an exponentially decaying behavior, suggesting that there should be a finite excitation gap, being well consistent with the result in the ground state, as the system in this case is almost dimerized;  for $J_{d}=1$, the specific heat exhibits a single peak, and when $T$ is very low, $C(T)$ obeys a polynomial behavior of the form $C(T)=1.741T-7.96T^2+22.51T^3-25.61T^4$. When $T\rightarrow 0$, $C(T)$ is linearly dependent on temperature, which indicates the existence of gapless excitations, and implies that the system is critical. It is also consistent with the preceding result that the ground state is an algebraic QSL.

Such criticality is further evidenced by the susceptibility at low temperature.The susceptibility is calculated according to $\chi(T)=[M(h+\Delta h)|_{T}-M(h)|_{T}]/\Delta h$, where $\Delta h=0.01$ is taken. The results for $J_{d}=10^{-4}$ and $1$ are presented in Fig.~\ref{fig13}. One may see that both curves obey the Curie-Weiss law at high temperature and exhibit a sharp peak at low temperature due to antiferromagnetic interactions. Significant differences occur when $T\rightarrow 0$. As shown in the inset of Fig.~\ref{fig13}, for $J_{d}=10^{-4}$, $\chi$ goes to zero in an exponential way,  revealing the existence of a finite spin gap, while for $J_{d}=1$, $\chi$ converges to a finite constant in a polynomial of the form $\chi(T)=0.09566-0.06828T+2.438T^2$, being reminiscent of a Luttinger liquid behavior, and consistent again with the critical feature of the ground state.

In addition, it is quite interesting to look at the Wilson ratio (WR) $R_w$ for the  present critical system at the isotropic point. The WR is defined by $R_w=(4/3)(\pi k_B/g\mu_B)^2 \chi/(C/T)$, where $\chi$ is the susceptibility, $C$ is the specific heat, $k_B$ is the Boltzmann constant, $g$ is the Land\'e factor, and  $\mu_B$ is the Bohr magneton. For simplicity we have assumed $k_B=g\mu_B=1$. It is known that for free electron gas, $R_w=1$. For most QSL theories, the WR is usually less than one. \cite{QSL} For the present system with $J_d=1$, at $T\rightarrow 0$, $\chi$ tends to a constant, and $C(T)\sim T$, which gives $R_w \approx 0.72$. In consideration of the fact that the linear temperature dependence of the specific heat resembles the Luttinger liquid behavior, and the WR $R_w$ is on the order of unity, which are analogous to the behaviors induced by fermionic quasiparticles, we conclude that the present isotropic system should be a fermionic gapless QSL.

\section{Conclusion}

The ground state and thermodynamic prosperities of the spin-$1/2$ HAFM on OKL have been systematically studied with the aid of powerful TN numerical simulations. We adopted three kinds of TN algorithms in calculations of the ground state energy per site, which gives $-0.4524$ by the infinite $D_c$ extrapolation in the thermodynamic limit, lower than $-0.4386(5)$ on kagom\'e lattice. The magnetic order is melted in the ground state due to strong frustration induced by corner sharing triangles. A QPT is found in this system. It is disclosed that below the QCP, the system has a finite spin gap and is in a VBS state, while above the QCP, the system is in a gapless QSL state. At the isotropic point, we uncover that the dimer-dimer correlation function decays algebraically, while the spin-spin and chiral-chiral correlation functions behavior in an exponential way. In addition, the specific heat at low temperature is shown to depend linearly on temperature, exhibiting a Luttinger liquid behavior, and the susceptibility tends to a finite constant when $T\rightarrow 0$, which indicates a gapless excitation in the system. The Wilson ratio is found to be 0.72, close to 1. All these features reveal that the isotropic spin-1/2 HAFM on OKL is a fermionic gapless QSL.

\acknowledgments

The authors acknowledge Wei Li, Xin Yan, and Yi-Zhen Huang for useful discussions, and also appreciate Guang-Zhao Qin and Xuan-Ting Ji for kind help. In particular, we thank Ying-Ying Tang and Zhang-Zhen He for useful discussions on the compounds with OKL. This work was supported in part by the MOST of China (Grants No. 2012CB932900 and No. 2013CB933401), the NSFC (Grant No. 14474279), and the Strategic Priority Research Program of the Chinese Academy of Sciences (Grant No. XDB07010100). SJR was supported by ERC ADG OSYRIS, Spanish MINECO (Severo Ochoa grant SEV-2015-0522, FOQUS grant FIS2013-46768), Catalan AGAUR SGR 874, Fundaci\'o Cellex, and EU FETPRO QUIC.

\end{document}